\newif\iflinenumbers
\linenumbersfalse 

\documentclass[10pt,oneside,notitlepage,abstracton,a4paper]{article}

\usepackage{epsfig} 
\usepackage{scrlayer-scrpage} 
\usepackage{amsmath} 
\usepackage{makecell}
\usepackage{graphicx}
\usepackage{booktabs}
\usepackage{arydshln}
\usepackage{setspace} 
\usepackage[utf8]{inputenc}
\usepackage[T1]{fontenc}

  \date{\normalsize \today}

\usepackage{chngcntr}

\usepackage[bottom]{footmisc}

\iflinenumbers
  \usepackage{lineno}
  \linenumbers
\fi

\usepackage[affil-it]{authblk}
 
\usepackage{etoolbox} 
\makeatletter 
\patchcmd{\@maketitle}{\LARGE \@title}{\fontsize{16}{19.2}\selectfont\@title}{}{}
\makeatother
 
\usepackage{geometry}
 
\newcommand{\imagepath}{./Plots}

\usepackage{xcolor}
\definecolor{DESYcyan}{RGB}{0,166,235}
\definecolor{DESYorange}{RGB}{242,142,0}
\definecolor{DESYgray}{RGB}{119,119,119}
\definecolor{bjetpurple}{RGB}{117,112,179}

\usepackage[compat=1.1.0]{tikz-feynman}
\usepackage{contour}
\usetikzlibrary{arrows,shapes,positioning}
\usetikzlibrary{decorations.markings}
\usetikzlibrary{decorations.pathmorphing}
\usetikzlibrary{decorations.pathreplacing}
\usetikzlibrary{patterns}
\usetikzlibrary{plotmarks}
\usetikzlibrary{shadows}

\usepackage[margin=8mm,font=small,labelfont=bf,format=plain]{caption}
\usepackage[margin=8mm,font=small,labelfont=bf,format=plain]{subcaption}

\usepackage{multirow} 

\newcommand{\subfigref}[1]{({\protect\subref{#1}})}

\usepackage[
  bookmarks,                   
  bookmarksopenlevel=1,        
  bookmarksnumbered=true,      
  pdfstartpage={1},            
  pdfstartview={FitH},         
  pdfkeywords={},              
  pdfsubject={},               
  pdfcreator={LaTeX with KOMA-Script and hyperref package},
  hyperfootnotes=true,         
  linkbordercolor={0 1 1},     
  menubordercolor={0 1 1},     
  urlbordercolor={1 0 0}       
]{hyperref}                    

\usepackage{cleveref}
 
\usepackage{slashed}

\newcount\colveccount
\newcommand*\colvec[1]{
        \global\colveccount#1
        \begin{pmatrix}
        \colvecnext
}
\def\colvecnext#1{
        #1
        \global\advance\colveccount-1
        \ifnum\colveccount>0
                \\
                \expandafter\colvecnext
        \else
                \end{pmatrix}
        \fi
}

\usepackage[english]{babel}
\usepackage{csquotes}

\newcommand{\Lumi}{\mathcal{L}}
\newcommand{\Pol}{\mathcal{P}} 
\newcommand{\PoleM}{\Pol_{\eM}}
\newcommand{\PoleP}{\Pol_{\eP}}

\newcommand{\eP}{e^{+}}
\newcommand{\eM}{e^{-}}

\newcommand{\muM}{\mu^{-}}

\newcommand{\invab}{\,\text{ab}^{-1}}

\renewcommand{\deg}{^{\circ}}

\newcommand{\Ae}{A_{e}}
\newcommand{\Af}{A_{\mu}}
\newcommand{\kR}{k_{R}}
\newcommand{\kL}{k_{L}}
\newcommand{\kN}{k_{0}}
\newcommand{\ef}{\epsilon_{\mu}}
\newcommand{\AFB}{A_{FB}}
\newcommand{\sigmaZero}{\sigma_{0}}

\title{Interplay of beam polarisation and systematic uncertainties at future e+e- colliders}
\author[1,2]{\underline{Jakob Beyer}\thanks{\textit{\footnotesize{Talk presented at}\\
\footnotesize{*** The European Physical Society Conference on High Energy Physics (EPS-HEP2021), ***}\\
\footnotesize{*** 26-30 July 2021 ***}\\
\footnotesize{*** Online conference, jointly organized by Universität Hamburg and the research center DESY ***}
}}}
\author[1]{Jenny List}
\affil[1]{Deutsches Elektronen-Synchrotron DESY, Notkestr. 85, 22607 Hamburg, Germany}
\affil[2]{Universit\"at Hamburg, Hamburg, Germany}

\begin{document}


\maketitle


\begin{abstract}
  Future high-energy $e^{+}e^{-}$ colliders will provide some of the most precise tests of the Standard Model. Statistical uncertainties on electroweak precision observables and triple gauge couplings are expected to improve by orders of magnitude over current measurements.

  This provides a new challenge in accurately assessing and minimising the impact of systematic uncertainties. Beam polarisation may hold a unique potential to isolate and determine the size of systematic effects. So far, studies have mainly focused on the statistical improvements from beam polarisation. This study aims to assess, for the first time, its impact on systematic uncertainties.

  A combined fit of precision observables, such as chiral fermion couplings and anomalous triple gauge couplings, together with experimental systematic effects is performed on generator-level differential distribution of 2-fermion and 4-fermion final-states. Different configurations of available beam polarisations and luminosities are tested with and without systematic effects, and will be discussed in the context of the existing projections on fermion and gauge boson couplings from detailed experimental simulations. 
\end{abstract}

\clearpage


\section{Introduction}

The latest update of the European strategy of particle physics identifies ``[a]n electron-positron Higgs factory [as] the highest-priority next collider''~\cite{European:2720129}.
Proposed future $\eP\eM$ Higgs factories have either no beam polarisation, $\eM$ beam polarisation or polarisation of both beams.
This situation raises the question which expected advantages beam polarisation can bring.
Among the frequently studied advantages are the increase in effective luminosity, the suppression of backgrounds and the analysis of chiral behaviour. In the context of dark matter searches, it has also been shown that the combination of datasets with different beam polarisations reduces the impact of global scaling systematic uncertainties drastically, if they are at least partially correlated between the data sets~\cite{Habermehl:2020njb}.
The study presented here investigates to which extent beam polarisation could help to control both global and shape-dependent systematic effects in the context of electroweak precision measurements. For this purpose a framework to simultaneously fit differential distributions from multiple physics processes and various data sets with different polarisations has been developed~\cite{beyer2021isolating}. The fit performs a Poissonian log-likelihood maximisation using the \texttt{Minuit2} framework of \texttt{ROOT}.

Proposed future colliders motivate the polarisation and luminosity configurations used in this study \cite{eppsupg2020physics}.
The polarisation scenarios are an unpolarised scenario ($\PoleM = \PoleP = 0\%$) with only one unpolarised dataset, a $\eM$-only polarised scenario with $\PoleM = \pm 80\%$ and $\PoleP=0\%$ and a luminosity sharing of $50:50$ between the two datasets, and a fully polarised scenario with $\PoleM = \pm 80\%$ and $\PoleP = \pm 30\%$ and an optimized luminosity sharing of $45:45:5:5$ where the first two are the opposite-sign and the second two the same-sign datasets.
Two luminosity scenarios with a factor 5 difference are tested: $\Lumi = 2\invab$ and $\Lumi = 10\invab$.

All polarisations and the luminosity can vary freely and can be given a Gaussian constraint to represent external measurements, e.g.\ here \cite{Vormwald:2015hla, Jelisav_i__2013}: 
$\Delta L / L = 3\times10^{-3}$ and $\Delta \Pol / \Pol = 2.5\times10^{-3}$ ($\Delta \Pol = 2.5\times10^{-3}$) for polarised (unpolarised) beams.
The helicity flip between polarised datasets is not assumed to be perfect, so that each signed polarisation is a separate parameter. 

As a prototype for shape-depending systematic effects, a simplified model of the muon acceptance is introduced as a parametrised detector effect as described in~\cite{beyer2021isolating}. The model assumes that muons are perfectly reconstructed in the central region down to a specific limiting angle below which they can not be reconstructed. The 
limiting angles can be varied independently for the forward and backward direction; their nominal values are chosen to be $7\deg$ here, inspired by the ILD detector \cite{theildcollaboration2020international}.

\section{Physical parameters and systematic effects}\label{subsec:PhysSysEffects}

This contribution focuses on di-muon production at $250\,$GeV, separated into events returning to the $Z$ pole and high-$\sqrt{s'}$ events, using generator-level events created for the $250\,$GeV ILD production~\cite{Berggren_2021}, produced with WHIZARD2.8~\cite{Kilian_2011, Moretti:2001zz}. Previous work used a set of processes with two and four fermions in the final state~\cite{Beyer:449759,beyer2021isolating}, but did not yet include the angular acceptance.

The di-muon distributions are one-dimensional distributions of the $\muM$ polar angle in the di-muon rest frame, separated into a high-energy part with $\sqrt{s'}\in[180,275]\,$GeV (``high-$Q^2$'') and radiative-return part with $\sqrt{s'}\in[81,101]\,$GeV (``return-to-Z'').

These distributions are described by six fit parameters for each of the two $\sqrt{s'}$ ranges. The first three are the 
unpolarised total cross-section $\sigmaZero$ as well as the initial- and final-state fermion asymmetries $\Ae$ and $\Af$,
as known from classic $Z$-pole physics, e.g.\ at LEP and SLC~\cite{Peskin:1988cz}. At $\sqrt{s}$ above the $Z$-pole, a fourth parameter $\ef$ is needed to model the effect of $Z/\gamma$ interference. Finally, two correction parameters $\kL$ and $\kR$ account for radiative corrections in the shape of the initial state radiation included in the Monte-Carlo samples.

In the case of an unpolarised collider, the $\Ae$, $\Af$ and $\ef$ parameters combine to $\AFB$, and  $\kL$ and $\kR$ merge into the unpolarised $\kN$. 
A measurement of all chiral parameters at an unpolarised collider would require the reconstruction of final state polarisations, i.e.\ from $\tau$-leptons. 

\section{Accessing chiral behavior with beam polarisation}
\begin{figure}
  \centering
  \includegraphics[width=0.75\textwidth]{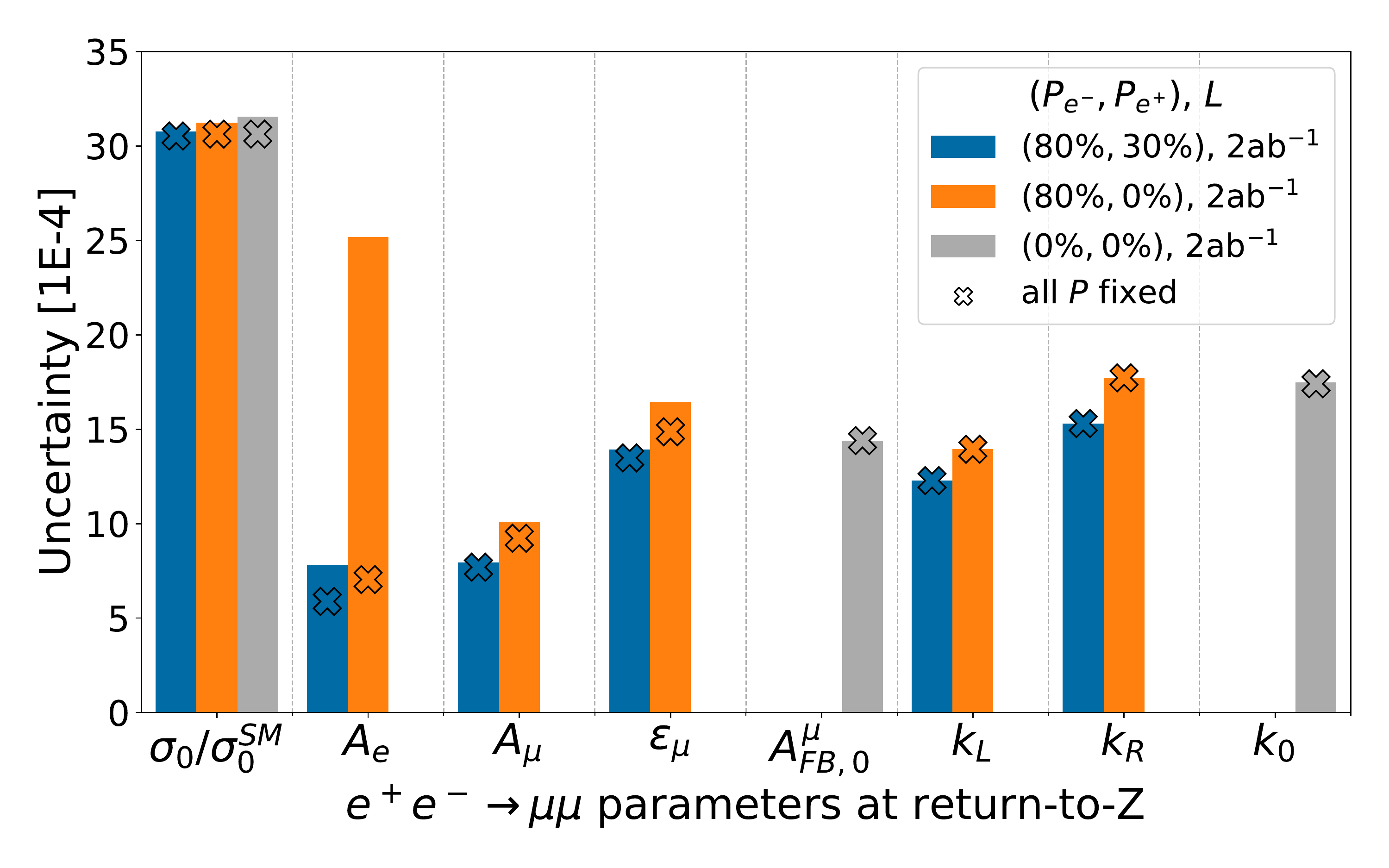}
  \caption{
    Precision on the return-to-Z muon pair production parameters for varius collider setups. 
    Crosses represent result with fixed polarisation amplitudes.
  }
  \label{fig:PhysImpact}
\end{figure}

Figure~\ref{fig:PhysImpact} shows the resulting precisions on the six (or three in the unpolarised case) fit parameters describing the $Z$-return samples for various assumptions on the availability of beam polarisation, all for an integrated luminosity of $2$\,ab$^{-1}$, and with fixed angular acceptance parameters.

The total unpolarised cross-section is determined in all cases at the level of $3\times 10^{-3}$, i.e.\ is -- in this setup -- only limited by the assumed precision of the luminosity measurement. In the case that both beams are polarised (blue), 
the two asymmetry parameters $\Ae$ and $\Af$ can be determined at the level of $7\times 10^{-4}$, which is about two times more precise than $\AFB$ in the unpolarised case (grey). This illustrate the significant additional information provided by the polarised data sets. It can also be noted that if both beams are polarised, there is hardly any additional uncertainty from the finite knowledge of the beam polarisation, which can be seen from the fact that there is no ($\Af$) or just a tiny ($\Ae$) improvement when the polarisation parameters are fixed in the fit (crosses). The situation changes drastically in the absence of positron polarisation (orange): in this case, $\Ae$ cannot be disentangled from the positron polarisation parameter and is limited directly by the assumed polarimeter precision --- in other words, a deviation from $P(e^+)=0$ cannot be distinguished from a non-SM value of $\Ae$. 

This can be seen more explicitly in Fig.~\ref{fig:hist_cor}, which shows the degree of correlation between all the parameters in the fit, in Fig.~\ref{fig:hist_cor_1pol} for the case of only the electron beam polarised, and in Fig.~\ref{fig:hist_cor_2pol} for the case of both beams polarised. The orange boxes highlight the correlation between
the polarisation parameters and $\Ae$ in the $Z$-return sample, clearly showing how a non-zero positron polarisation reduces the correlations. The same effect can also be seen for $\Ae$ in the high-$Q^2$ sample. On a more subtle level, also correlations between the radiative correction parameters on one hand and the total cross section and the $\Ae$ parameters on the other hand are reduced if both beams are polarised.  

\newcommand{\halffraction}{0.49}
\begin{figure}
  \centering
  \begin{subfigure}[t]{\halffraction\textwidth}
    \centering
    \includegraphics[width=\textwidth,trim={1.5cm 1.5cm 1.5cm 1.5cm},clip]{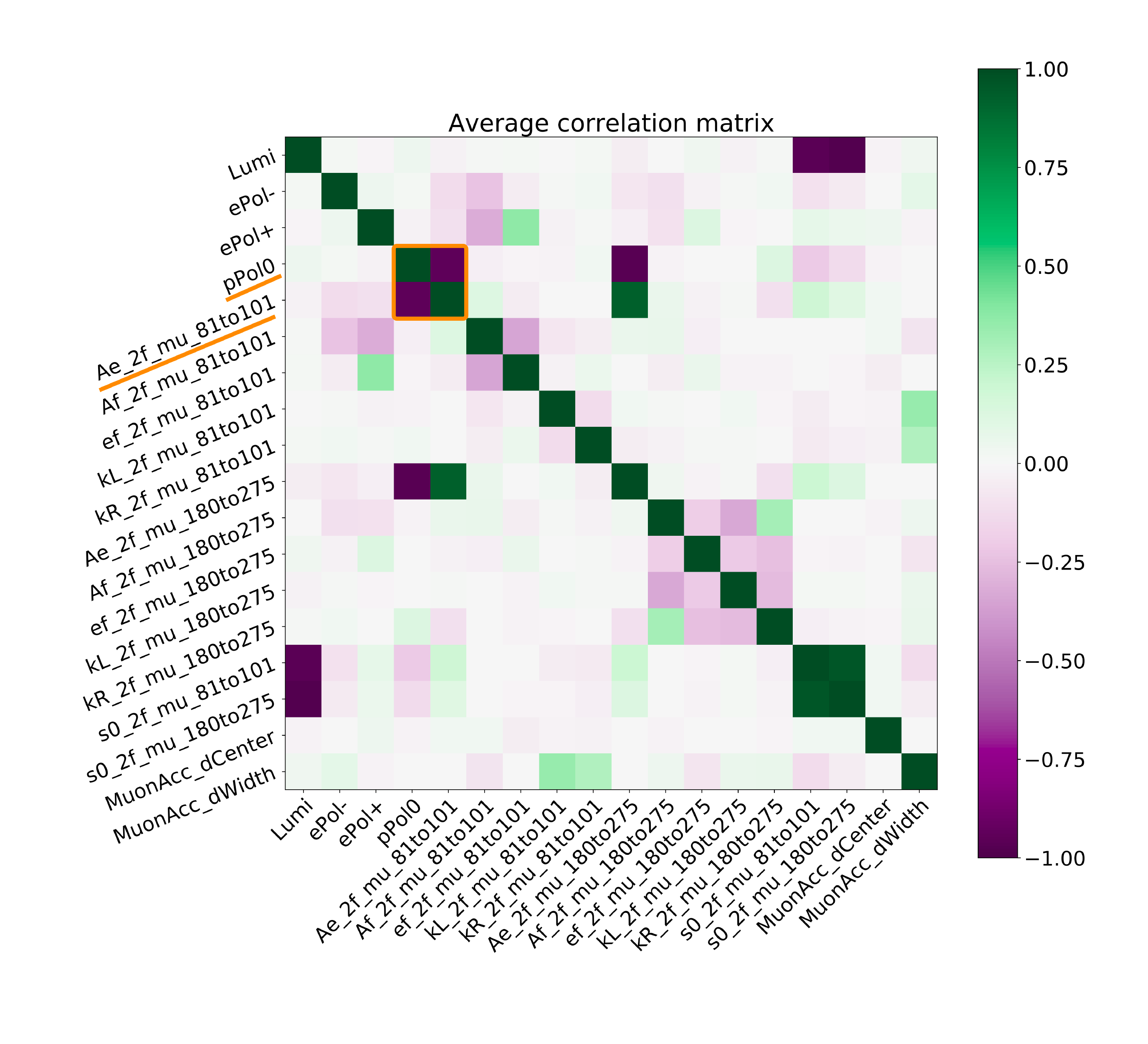}
    \caption{}\label{fig:hist_cor_1pol}
  \end{subfigure}
  \begin{subfigure}[t]{\halffraction\textwidth}
    \centering
    \includegraphics[width=\textwidth,trim={1.5cm 1.5cm 1.5cm 1.5cm},clip]{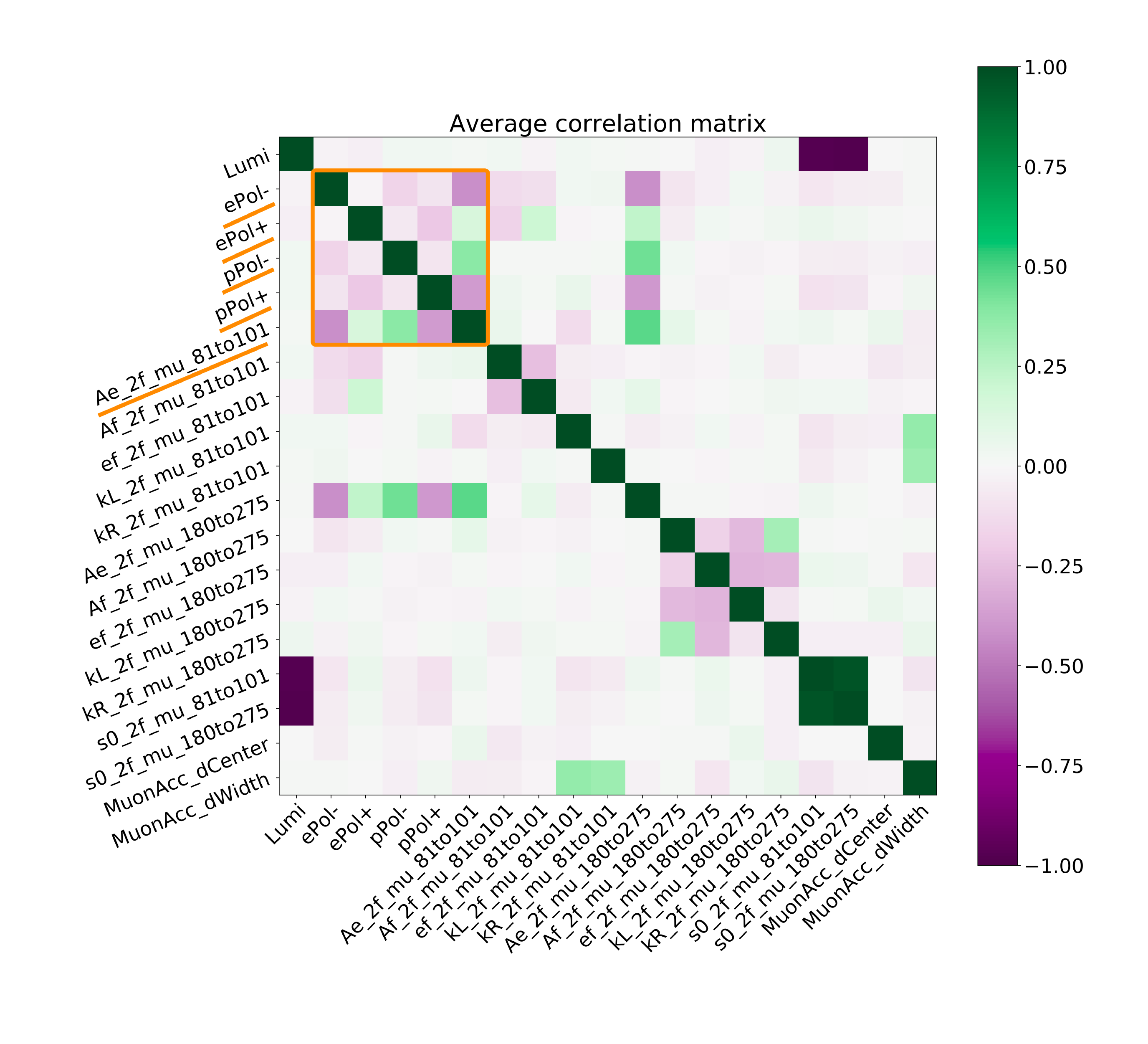}
    \caption{}\label{fig:hist_cor_2pol}
  \end{subfigure}%
  \caption{%
    Correlation matrices of all parameters included in the fit for $\Lumi = 2\invab$ and scenarios of: \subfigref{fig:hist_cor_1pol} only $\eM$ polarisation, \subfigref{fig:hist_cor_2pol} both beams polarised.
    The correlations between the return-to-Z $\Ae$ parameter and the polarisation amplitudes are emphasized.
  }
  \label{fig:hist_cor}
\end{figure}

Once the 4-fermion processes will be included, there will be further contraints on the polarisation amplitudes, so that the polarisation will be known better than just the polarimeter precision~\cite{Karl:424633}. Still, the full correlation between $\Ae$ and $\PoleP$ remains, propagating any potential bias of $\PoleP$ directly to $\Ae$. For the same reason, a contribution of the di-fermion final states to the polarisation measurement is only possible if both beams are polarised.

\section{Isolating detector effects with beam polarisation}
The notion of separating effects by their chiral behavior can be translated to the interplay of physical and systematic effects. Detector effects are uniquely chirality-independent, compared to their generally chirality-independent physical counterparts. A combined fit to all polarised datasets of a polarised collider could use the different chiral behaviour to reduce correlation between physical and systematic effects. This may reduce the impact of systematic uncertainties (e.g from detector effects) which can be assumed to be correlated between data sets with different polarisation signs. This is
usually the case when the beam helicities can be flipped sufficiently fast, e.g.\ at the level of a few Hz, and in a randomised way. This has been shown for global scaling uncertainties, like the luminosity measurement~\cite{Habermehl:2020njb}, and will be investigated in this study with a shape-dependent effect, taking as example the angular acceptance for muon reconstruction introduced above.

\begin{figure}
  \centering
  \includegraphics[width=0.5\textwidth]{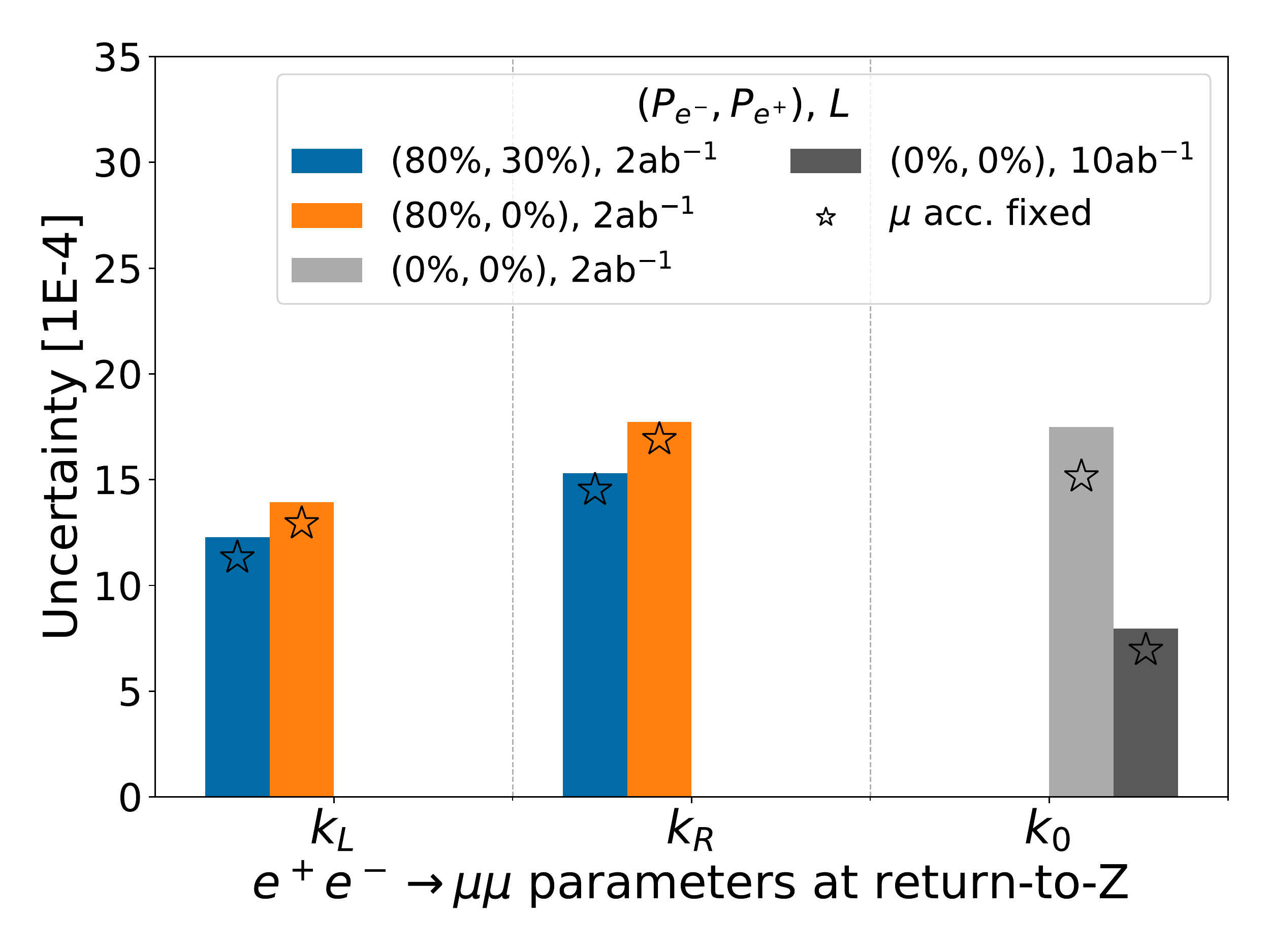}
  \caption{%
    Precision on the radiative correction parameters for return-to-Z muon pair production for varius collider setups. 
    Stars represent result with fixed muon acceptance parameters.
  }
  \label{fig:SystImpact}
\end{figure}

In this very simplified case of just two parameters describing a perfect ``box-like'' acceptance, it turns out that the only change to the results presented in the previous section occurs for the radiative-correction parameters $k$. Figure~\ref{fig:SystImpact} compares the precisions on these parameters for the case of fixed muon acceptance (stars) with the case of two additional free acceptance parameters.
The additional systematic uncertainty due to the muon acceptance makes up $14\%$ of the total uncertainty on the $k$ parameters in the unpolarised case.
Adding $\eM$ beam polarisation leads to a relative impact of $7(5)\%$ on individual $\kL(\kR)$ parameters.
An additional polarisation of the $\eP$ does not lead to a further reduction of the relativ systematic uncertainty.\footnote{Subsequent studies have shown that this a consequence of the choice of parameters which lead to correlations between $\kL$ and $\kR$. The same $14\%$ impact is observed in the polarised cases when choosing a more appropriate linear combination of $\kL$ and $\kR$.}

In case of the di-muon events, small variations the perfect box-acceptance around the default $7^\circ$ in the detector system thus lead to very limited changes in the angular distributions in the restframe on the di-muon, which have very different shapes that the the effects of the other parameters and thus can be easily determined from the data themselves.
Even if so far only tested in this very simplified Ansatz, this robustness of the physics parameters w.r.t.\ the acceptance is in principle very good news.   

In remains to be studied in the future, however, whether this statement holds up also for the four-fermion case, with more subtle angular effects e.g.\ from anomlous triple gauge couplings, and for more realistic parametrisations of the acceptance. 

\section{Conclusion}
Beam polarisation separates effects by their chiral behaviour.
This is especially relevant for those effects with the same or a similar differential shape.
This chiral sensitivity is essential to gain direct access to all chiral parameters in muon pair production - and difermion production in general.
A single polarised beam gives access to the full set of chiral parameters.
Polarising both beams also removes remaining full correlations between the chiral parameters and the polarisation amplitudes. In general, every additional beam polarisation decreases correlations between parameters.
A first, very simplified Ansatz to include a systematic uncertainty on the muon acceptance profed the fit results to be very robust against the exact knowledge of the onset of the acceptance in the forward (and backward) region of the detector. In the future it needs to be investigated in how far this result holds also in the case of more realistic parametrisations.


\subsection*{Acknowledgments}
This work was funded by the Deutsche Forschungsgemeinschaft under Germany’s Excellence Strategy – EXC 2121 ``Quantum Universe'' – 390833306.
It has benefited from the computing services provided by the German National Analysis Facility (NAF)\cite{Haupt_2010}.


\bibliographystyle{JHEP}


\end{document}